\documentclass[pdflatex,sn-mathphys-num]{sn-jnl}

\usepackage{graphicx}%
\usepackage{multirow}%
\usepackage{amsmath,amssymb,amsfonts}%
\usepackage{amsthm}%
\usepackage{mathrsfs}%
\usepackage[title]{appendix}%
\usepackage{xcolor}%
\usepackage{textcomp}%
\usepackage{manyfoot}%
\usepackage{booktabs}%
\usepackage{algorithm}%
\usepackage{algorithmicx}%
\usepackage{algpseudocode}%
\usepackage{listings}%
\usepackage{hyperref}
\definecolor{myred}{RGB}{255,0,0}

\theoremstyle{thmstyleone}%
\theoremstyle{thmstyletwo}%

\theoremstyle{thmstylethree}%

\begin{document}

\title[Article Title]{XDXD: End-to-end crystal structure determination with low resolution X-ray diffraction.}


\author[1]{\fnm{Jiale} \sur{Zhao}}

\author[2]{\fnm{Cong} \sur{Liu}}

\author[6]{\fnm{Yuxuan} \sur{Zhang}}

\author[3]{\fnm{Chengyue} \sur{Gong}}

\author*[4]{\fnm{Zhenyi} \sur{Zhang}}\email{zhenyi.zhang@bruker.com}

\author*[5]{\fnm{Shifeng} \sur{Jin}}\email{shifengjin@iphy.ac.cn}

\author*[6]{\fnm{Zhenyu} \sur{Liu}}\email{zyliu06@gmail.com}

\affil[1]{\orgdiv{Key Laboratory of Intelligent Information Processing of Chinese Academy of Sciences (CAS)}, \orgname{Institute of Computing Technology}, \country{China}}

\affil[2]{\orgdiv{AMLab},\orgname{University of Amsterdam}, \country{the Netherlands}}

\affil[3]{\orgname{University of Texas at Austin}, \country{the United States}}

\affil[4]{\orgname{Bruker Scientific Instruments (shanghai) Co., Ltd}, \country{China}}

\affil[5]{\orgname{Institute of Physics CAS}, \country{China}}

\affil[6]{\orgname{Independent Researcher}}


\abstract{
Determining crystal structures from X-ray diffraction data is fundamental across diverse scientific fields, yet remains a significant challenge when data is limited to low resolution. While recent deep learning models have made breakthroughs in solving the crystallographic phase problem, the resulting low-resolution electron density maps are often ambiguous and difficult to interpret. To overcome this critical bottleneck, we introduce XDXD, to our knowledge, the first end-to-end deep learning framework to determine a complete atomic model directly from low-resolution single-crystal X-ray diffraction data. Our diffusion-based generative model bypasses the need for manual map interpretation, producing chemically plausible crystal structures conditioned on the diffraction pattern. We demonstrate that XDXD achieves a 70.4\% match rate for structures with data limited to 2.0~\AA{} resolution, with a root-mean-square error (RMSE) below 0.05. Evaluated on a benchmark of 24,000 experimental structures, our model proves to be robust and accurate. Furthermore, a case study on small peptides highlights the model's potential for extension to more complex systems, paving the way for automated structure solution in previously intractable cases.

}




\maketitle

\section{Introduction}\label{sec1}

The three-dimensional architecture of molecules and materials dictates their function, and for over a century, X-ray crystallography has been the definitive method for its determination\citep{xie2018crystal,jumper2021highly}. The central challenge remains the crystallographic phase problem: diffraction experiments measure structure factor amplitudes but lose phase information. Historically, early methods relied on laborious trial-and-error computations, with the Patterson method\cite{rius2014application} and computational aids like Beevers-Lipson strips\cite{beevers1936numerical} providing incremental advances. A key advance came in the 1950s–60s when Karle and Hauptmann established direct methods\cite{hauptman1954solution,sayre1952squaring}, building on Sayre's foundational equations\cite{giacovazzo1998direct,hauptman1953solution} provides probabilistic solutions to the phase problem using intensity statistics. However, these traditionally require high-resolution diffraction data (typically better than 1.2~\AA{}), creating barriers for weakly diffracting crystals like complex biological samples or materials under non-ambient conditions.


While molecular replacement\cite{gorelik2023molecular} and recent deep learning approaches like PhAI\cite{larsen2024phai} have attempted to address low-resolution challenges, significant limitations persist. PhAI demonstrated phase recovery at 2.0~\AA{} resolution—proving sufficient information exists in low-resolution data—but remains restricted to a few centrosymmetric space groups (limiting phases to 0 or $\pi$). Critically, its output is an electron density map that, at low resolution, lacks clear atomic features. This renders subsequent accurate atomic model building subjective, time-consuming, and often intractable for human crystallographers. Meanwhile, existing crystal structure prediction models still struggle with systems exceeding 52 atoms on simple dataset such as Perov-5, MP-20 and ICSD\cite{zagorac2019recent,castelli2012new,castelli2012computational,jain2013commentary,zagorac2019recent,zeni2025generative, jiao2023crystal}, and powder diffraction methods\cite{riesel2024crystal, lai2025end, guo2025ab, guo2024towards, chen2024crystal,li2024powder} suffer from information loss in 1D signals.

To bridge this crucial gap between phasing and model building, we propose XDXD (X-ray Diffusion for structure Determination), the first end-to-end generative model {to our knowledge} that predicts a complete crystal structure directly from low-resolution single-crystal diffraction data. Our approach is inspired by recent successes of generative models in solving crystal structures from powder X-ray diffraction (PXRD) data\cite{riesel2024crystal, lai2025end, guo2025ab, guo2024towards, chen2024crystal,li2024powder}, which have proven effective at generating chemically rational structures. XDXD leverages a diffusion-based framework that is conditioned on the experimental diffraction amplitudes to generate a full set of atomic coordinates, thereby bypassing the ambiguous process of map interpretation entirely.

We demonstrate that our model, trained on a diverse set of 395,117 simulated diffraction pattern, can successfully determine atomic models from experimental data limited to 2.0~\AA{} resolution. Based on advanced model designing, it handles unit cells containing 0–200 non-hydrogen atoms, far exceeding prior limitations. We validate XDXD on a large-scale test set of approximately 24,000 experimental structures from the Crystallography Open Database (COD)\cite {gravzulis2009crystallography,downs2003american}, covering a wide range of space groups and chemical compositions. By generating multiple candidate structures and ranking them against the experimental data, our workflow robustly identifies the correct atomic model. Notably, XDXD shows the ability to determine peptide structures without being trained on such data, suggesting its potential for proteins, nucleic acids, and complexes. This work represents a significant step towards a fully automated pipeline for crystal structure determination, promising to unlock structural insights from a vast number of challenging cases previously limited by data resolution.

\section{Results}\label{sec2}

\subsection{Overview of XDXD model.}\label{subsec}
\begin{figure}[H]
\centering
\includegraphics[width=1\textwidth]{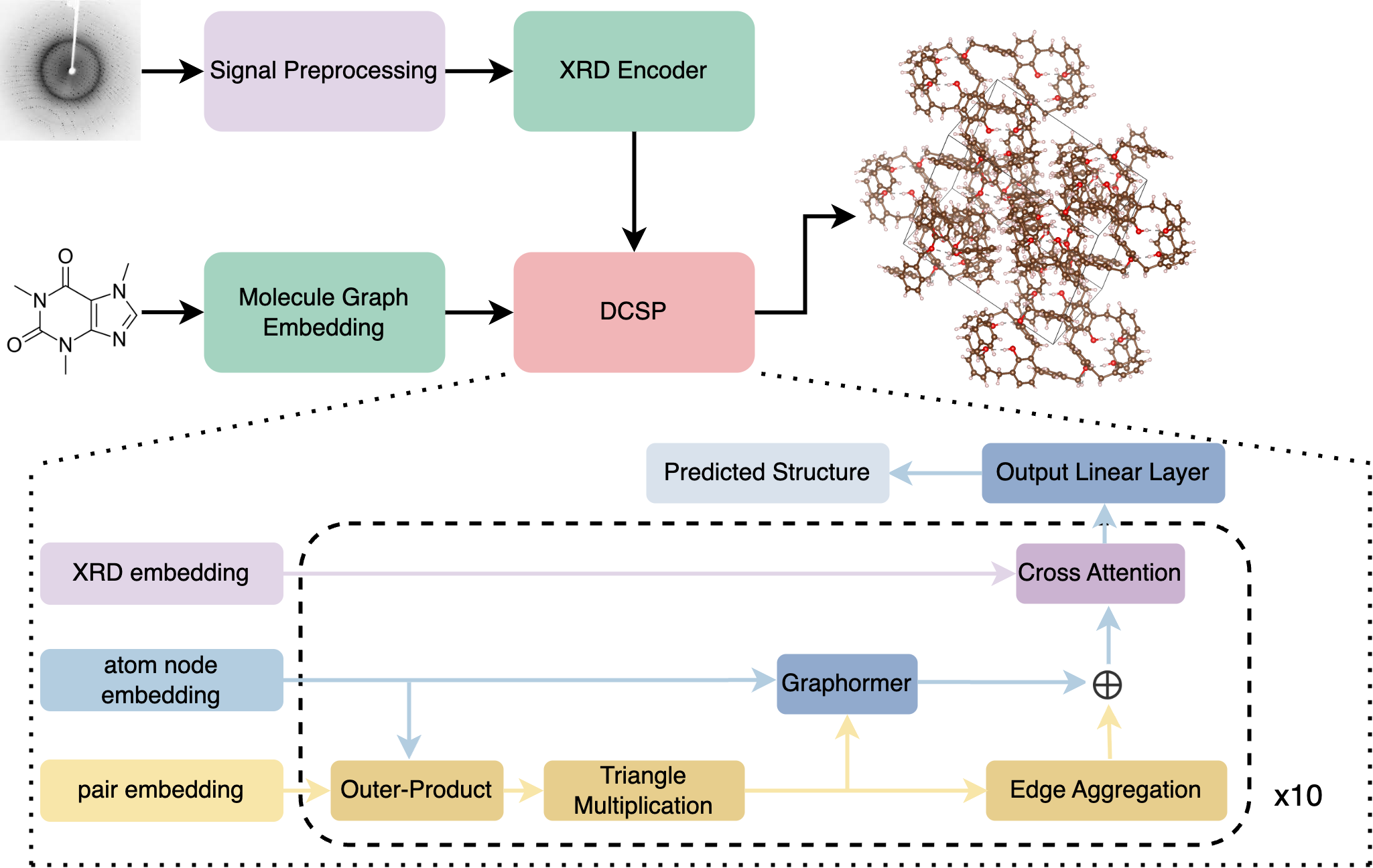}
\caption{
\textbf{XDXD model architecture.} 
The X-ray diffraction (XRD) signal is first preprocessed and then passed through an XRD encoder to obtain the corresponding XRD embedding. In parallel, the molecular graph embedding layer encodes atom types and bonds. Subsequently, the embeddings from the XRD encoder and the molecular graph embedding layer are jointly fed into the Diffraction Signal Conditioned Crystal Structure Prediction (DCSP) module. Within the DCSP module, pair embeddings are computed via outer-product, tri-multiplication, Graphormer layers, and edge aggregation operations. The final structure prediction is derived from the atom node embeddings. Meanwhile, atom node embeddings interact with pair embeddings through outer-product operations and Graphormer modules. Additionally, the XRD embeddings and atom node embeddings are integrated through a cross-attention mechanism.
}\label{model}
\end{figure}

The XDXD model is an end-to-end deep learning framework designed to predict a complete atomic crystal structure directly from a given chemical composition and its corresponding single-crystal X-ray diffraction (XRD) signal. As illustrated in \textbf{Fig.~\ref{model}}, the architecture of XDXD consists of three main components. First, an XRD Encoder, constituted with transformer layers, processes the pre-processed diffraction signal to produce X-ray diffraction signal embeddings that captures the global information from all reflections. In parallel, a Molecular Graph embedding layer encodes the system's chemical information, such as atom types. These two embeddings are then fed into the central Diffraction-Conditioned Structure Predictor (DCSP) module, a diffusion-based generative model that iteratively refines atomic coordinates to produce the final structure. {The overall structure solution workflow begins by generating a set of 16 candidate structures, each initiated from random noise.} For each generated candidate, a theoretical diffraction pattern is simulated and compared against the experimental input data. The candidates are then ranked based on the cosine similarity score between their simulated pattern and the experimental pattern, and the top-ranked structure is reported as the final prediction. The workflow of our crystal structure prediction is shown in \textbf{Supplementary Fig.~\ref{workflow}}.

Theoretically, a crystal's electron density $\rho(\mathbf{r})$ can be derived from its complex-valued structure factors $F(\mathbf{h})$ through the inverse Fourier transform:

\begin{equation}
\rho(\mathbf{r}) = \frac{1}{V} \sum_{\mathbf{h}}  e^{-2\pi i \mathbf{h} \cdot \mathbf{r} }F(\mathbf{h})
\end{equation}

The equation is analogous to the cross-attention mechanism shown below. Specifically, the term $\frac{\mathbf{Q}^{l-1} \mathbf{K}^T}{\sqrt{d_k}}$ corresponds to $e^{-2\pi i \mathbf{h} \cdot \mathbf{r}}$, while $F(\mathbf{h})$ corresponds to the value matrix, $\mathbf{V}$. Consequently, we employ cross-attention as the conditioning mechanism in our model.
\begin{equation}
\mathbf{a}^{l} = \mathbf{a}^{l-1} + \text{softmax}\left( \frac{\mathbf{Q}^{l-1} \mathbf{K}^T}{\sqrt{d_k}} \right) \mathbf{V}
\end{equation}
Here, $a^l$ denotes atom representations at layer $l$, while $Q$, $K$, $V$ represent query, key, and value matrices respectively. To simulate experimental noise conditions – particularly elevated uncertainties in observed structure factors ($\sigma(F_{\mathrm{o}})$) – we introduce random signal dropout during training. For each sample, we uniformly select a drop ratio between 0-10\% and randomly remove corresponding diffraction signals.

Due to the translational invariance of crystals, we embed only the relative vector between two atoms in the edge embedding, rather than their coordinates. This makes edge embeddings crucial for structure prediction. During the development of our model, we observed instability and inefficiency during training, which we attribute to insufficient interaction between node and edge representations. In the baseline architecture, critical three-dimensional information is primarily encoded within the edge features. This spatial context is then only indirectly propagated to the nodes, for instance, through a Graphormer-style\cite{ying2021transformers} bias term. We argue that this limited, one-way communication channel creates a bottleneck, preventing the model from fully leveraging geometric information.
To address this issue, we introduce a set of modules designed to foster a more direct and bidirectional flow of information. 
First, an Edge Aggregation Module facilitates an edge-to-node update, enriching each node's representation by directly incorporating the features of all its incoming edges. This allows nodes to assimilate comprehensive relational context. Concurrently, an Outer-Product Module enables the reciprocal node-to-edge update, dynamically informing edge representations by aggregating the embeddings of their connected nodes. Finally, to enforce geometric consistency and capture higher-order relational patterns, we employ Triangle Multiplication on the edge representations, reinforcing the model’s spatial awareness across relational triplets.
Further details are provided in the Methods section.

\subsection{Overall performance on experimental data}\label{subsec}



\begin{figure}[htbp]

\centering
\includegraphics[width=1\textwidth]{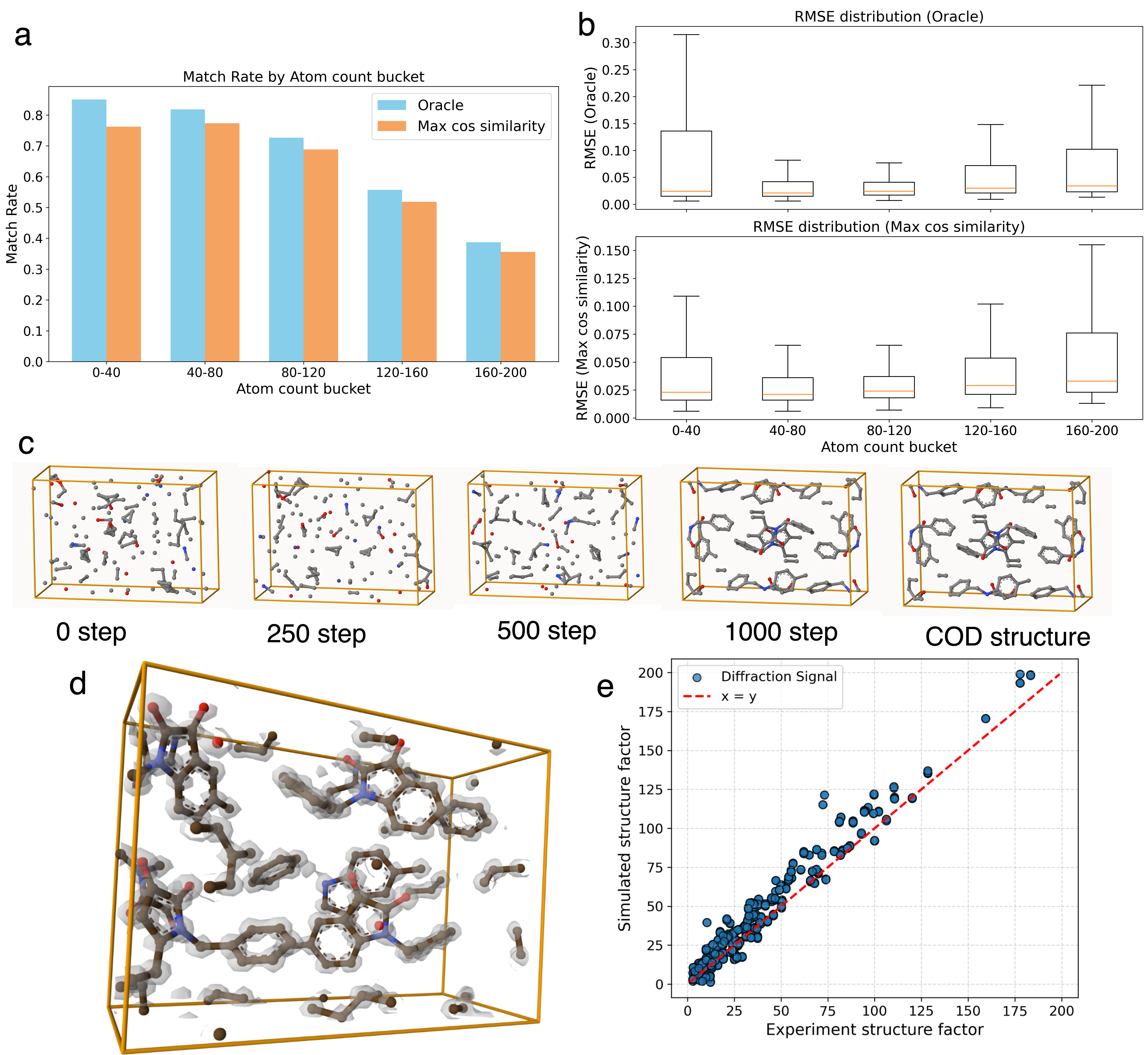}
\caption{
\textbf{Evaluation results on experimental data.}  
\textbf{a.} The match rate is analyzed across various atom-count intervals.
\textbf{b.} The Root Mean Square Error (RMSE) is evaluated across different atom-count intervals.
\textbf{c.} The generation process for C\textsubscript{16}H\textsubscript{13}NO\textsubscript{2} with the space group `\textit{P b c a}'.
\textbf{d.} Electron density is simulated using the experimental amplitude and a simulated phase derived from the predicted structure.
\textbf{e.} The experimental structure is compared with the structure factor simulated from the predicted structure.
}\label{fig2}
\end{figure}

We evaluated our model on 24,000 experimental structures determined by X-ray diffraction from the COD dataset, retaining only diffraction signals with a resolution better than 2.0~\AA{}. Benefiting from the diffusion methodology, our model is able to generate diverse crystal structures by introducing different atomic positions initialized with uniform distribution. We employ two evaluation strategies: oracle evaluation and maximum cosine similarity ranking. For oracle evaluation, we select the structure with the lowest RMSE as the final reported result. This reflects the upper-limit performance of XDXD and allows for comparison with the cosine similarity ranking method to demonstrate the effectiveness of the ranking approach. However, oracle {evaluation} does not reflect realistic scenarios, as the ground truth structure is typically unknown. In practice, since we have both the experimental diffraction signals and the predicted structures, we can simulate the diffraction signals from the predicted structures and compute the cosine similarity between the simulated and experimental signals. This cosine similarity serves as a scoring metric to rank the predicted structures, allowing us to select and report the structure with the highest cosine similarity.

We evaluate the predicted structures using match rate and RMSE, which are calculated with the \texttt{pymatgen} library as detailed in the Methods section. As shown in \textbf{Fig.~\ref{fig2}a}, the match rate decreases as the number of atoms in the unit cell increases. 
Notably, the match rate still reaches around 40\% even for systems with 160–200 atoms, which is a strong result at this scale. As shown in \textbf{Fig.~\ref{fig2}b}, the RMSE of our model increases as the number of atoms in the structure rises. Although the mean RMSE for the 0–40 atom count range is relatively low, its upper quartile exceeds 0.1. This disparity arises because the RMSE is calculated only for matched structures. For few atom structures, it is comparatively easier to obtain a match, even when the predicted positions of some atoms deviate significantly from their actual positions.

To demonstrate the generative process, we detail an example of predicting an experimental structure in \textbf{Fig.~\ref{fig2}c}. The process begins with a structure initialized with uniform random noise. Our model, XDXD, then conditions on the experimental diffraction pattern to gradually and accurately determine the final crystal structure. {Electron density was calculated using the experimental intensities combined with phases calculated from our predicted structure.} The corresponding electron density map and refined crystal structure, shown in \textbf{Fig.~\ref{fig2}d}, exhibit that structure predicted by XDXD extremely high consistency with the experimental data. Finally, \textbf{Fig.~\ref{fig2}e} shows the correlation between the experimental and simulated structure factors, revealing a strong linear relationship that further validates the reliability and accuracy of our predicted structure.

\subsection{Consistency between experimental data and predicted structure.}\label{subsec}
\begin{figure}[htbp]
\centering
\includegraphics[width=1\textwidth]{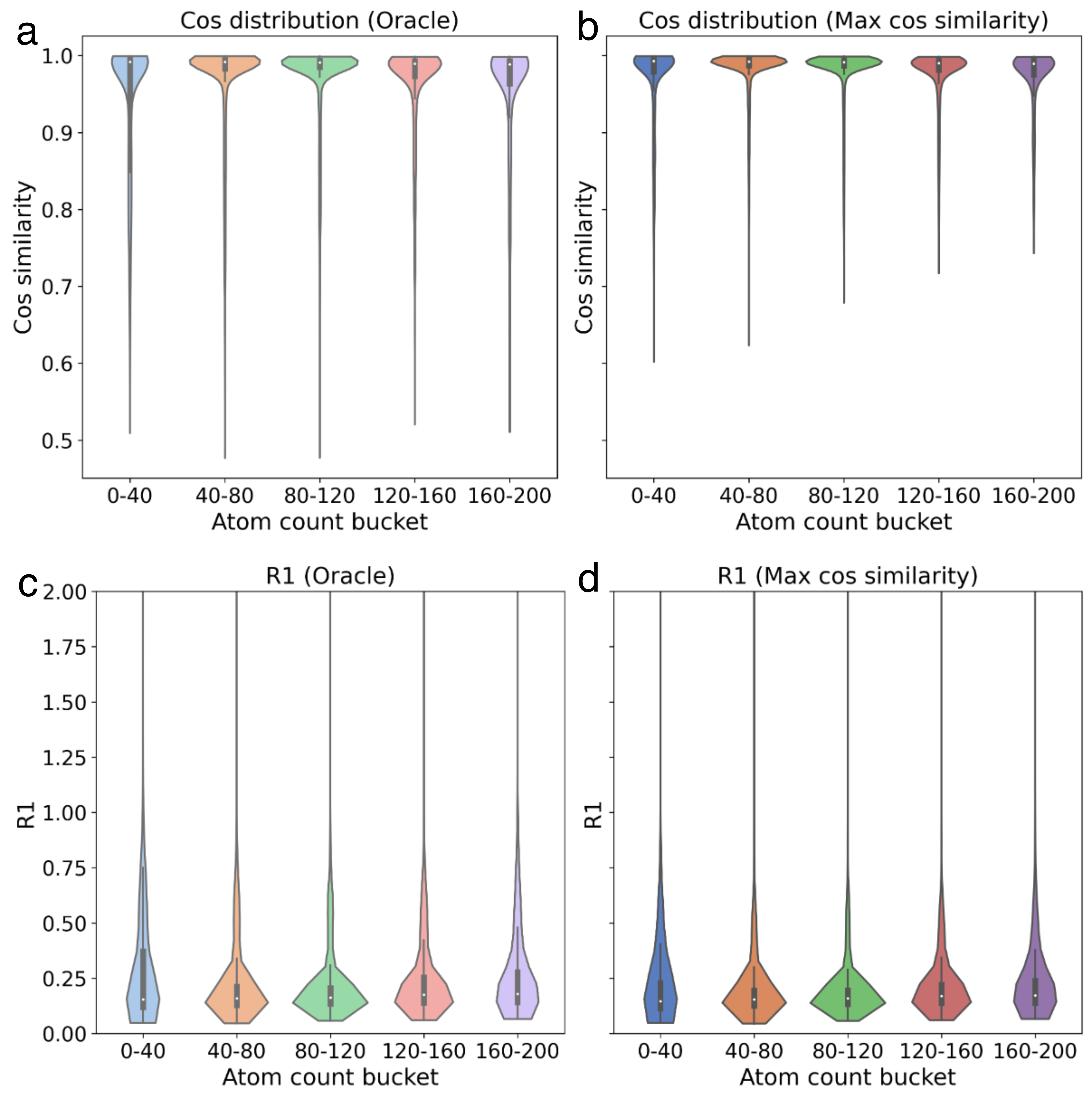}
\caption{
\textbf{Evaluation of diffraction pattern consistency.}
\textbf{a, b.} Cosine similarity between structure factors simulated from the predicted structure and the experimental structure, evaluated for signal resolutions lower than 2~\AA.  
\textbf{c, d.} R1 value between structure factors simulated from the predicted structure and the experimental structure, for signal resolutions below 2~\AA.
}\label{fig3}
\end{figure}

To further validate model performance, we computed cosine similarity between experimental diffraction patterns and those simulated from our predicted structures (\textbf{Fig.~\ref{fig3}a,b}). Our analysis reveals that, as the number of atoms increases, the overall cosine similarity tends to decrease—a trend that is expected, since the challenge of structure prediction becomes significantly greater for larger systems. Additionally, we calculated the R1-factor between the experimental and simulated diffraction patterns (\textbf{Fig.~\ref{fig3}c,d}). The consistently low values observed in this analysis further validate the reliability of our model and are in good agreement with the low RMSE shown in \textbf{Fig.~\ref{fig2}b}.

Experimental diffraction data inevitably contains resolution-dependent noise, leading to relatively lower signal-to-noise ratios (\(S/N\)) in low-resolution regions. However, the limited number of independent observations in low-resolution shells also means the number of variables can easily exceed the number of independent experimental observations, creating a significant risk of overfitting to the diffraction data \cite{brunger1992free}.
We also need to test for potential overfitting to the 2~\AA{} resolution shell, as this could artificially inflate the model's performance metrics. To do so, we recalculated the cosine similarity and R1-factor using only those reflections with a resolution higher than 2~\AA{} that satisfied the condition \(F_\mathrm{o} > 4\sigma(F_\mathrm{o})\).
As shown in \textbf{Supplementary Fig.~\ref{cos_all}} and \textbf{Supplementary Fig.~\ref{r1_all}}, the agreement is only slightly reduced. 
Very few cases shows R1-factor greater than 0.75 suggesting that cases of overfitting to signals at resolutions better than \(2\,\text{\AA}\) remain rare since there are very few cases with R1 greater than 0.75.

\subsection{Evaluation on different space group and element composition.}\label{subsec}
\begin{figure}[H]
\centering
\includegraphics[width=1\textwidth]{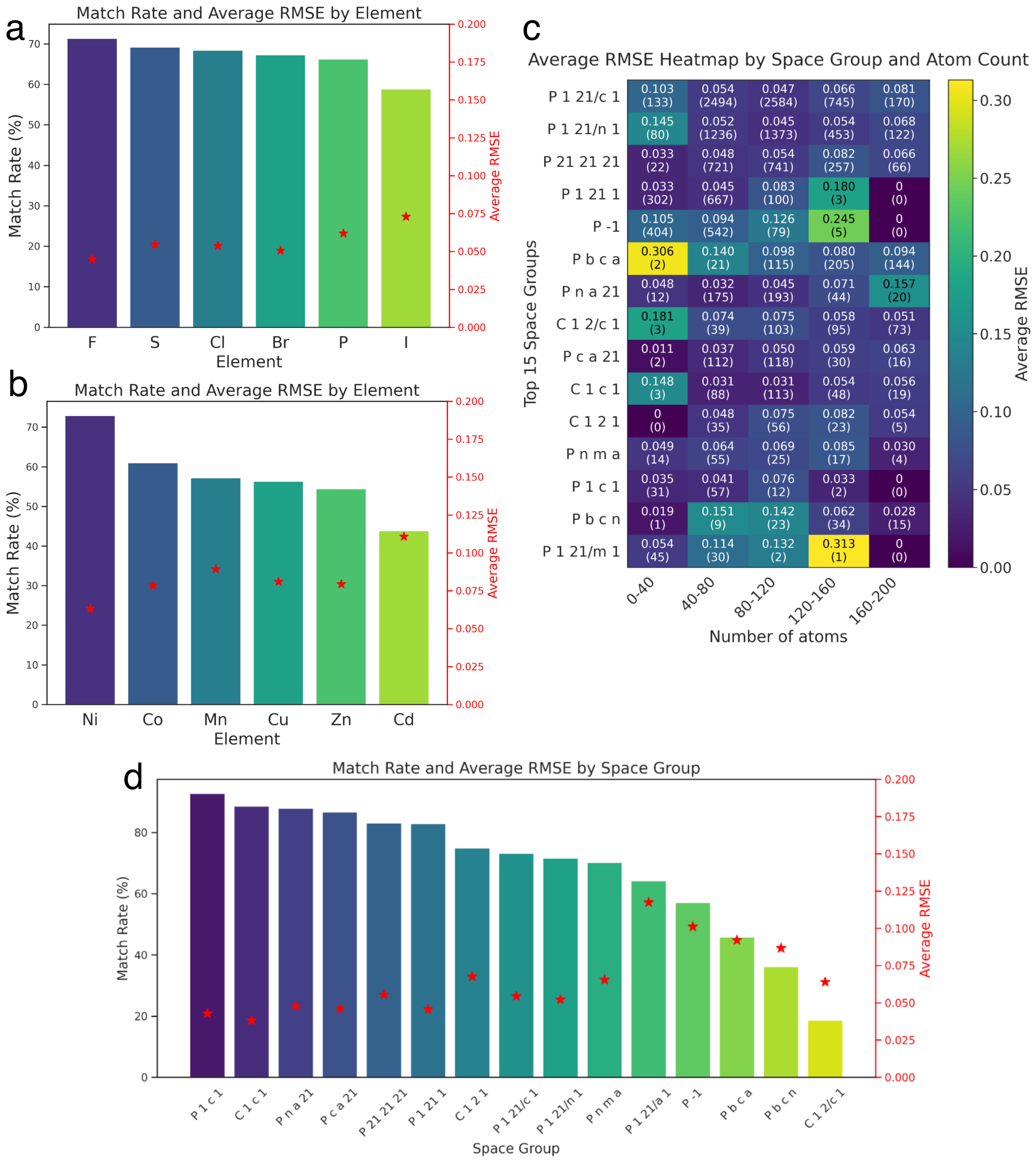}
\caption{
\textbf{Evaluation Results Based on Space Group and Elemental Composition.}  
\textbf{a.} Evaluation of match rate and RMSE for crystals containing only non-metal elements.  
\textbf{b.} Evaluation of match rate and RMSE for crystals containing metal elements.  
\textbf{c.} Heatmap of RMSE as a function of the number of atoms and space groups. The number in parentheses under RMSE denotes the number of samples in the test set.
\textbf{d.} Evaluation of match rate and RMSE across different space groups.
}\label{fig4}
\end{figure}

Solving crystal structures requires handling the complexities raised by different space group symmetry and elemental composition—a process demanding robust performance across both dimensions. We further evaluated the performance of XDXD on various elements and space groups. 

The element composition distributions for the training set and the COD test set are presented in \textbf{Supplementary Fig.~\ref{train_ele}} and \textbf{Supplementary Fig.~\ref{test_ele}}, respectively. We show the performance of XDXD on structures containing the six most common metal and non-metal elements, as shown in \textbf{Fig.~\ref{fig4}a} and \textbf{Fig.~\ref{fig4}b}. The solution process reveals consistently higher match rates and lower RMSE for non-metal systems, while atom-count-dependent RMSE in \textbf{Fig.~\ref{fig4}b} demonstrates scalability across system sizes. 

The space group distribution in the training set is shown in \textbf{Supplementary Fig.~\ref{train_sg}}, while the distribution for the COD test set is provided in \textbf{Supplementary Fig.~\ref{test_sg}}. XDXD was trained on 143 space groups. As shown in \textbf{Fig.~\ref{fig4}d,e}, the model maintains fidelity across 15 representative space groups and varying atom counts, confirming adaptability throughout the solution process.


\subsection{Effectiveness of diffraction pattern for XDXD.}\label{subsec}

The critical dependence of structure determination performance on diffraction signal quality was quantified through systematic ablation studies (\textbf{Supplementary Fig.~\ref{abl_res}}). Incorporation of X-ray diffraction data substantially enhances predictive accuracy: When diffraction signals are excluded entirely, the single-shot match rate on the validation set plateaus at merely 34\%. Introduction of diffraction data at 3.0 \AA{} resolution elevates accuracy to 53\%, while refinement to 1.5 \AA{} resolution further increases the match rate to 65\%. This progressive enhancement demonstrates that higher-resolution diffraction data - containing richer reciprocal-space information - directly and positively correlates with determination fidelity. Experimental protocols and computational parameters governing these investigations are comprehensively documented in the Methods section.

\subsection{Experimental validation.}\label{subsec}



\begin{figure}[H]
\centering
\includegraphics[width=1\textwidth]{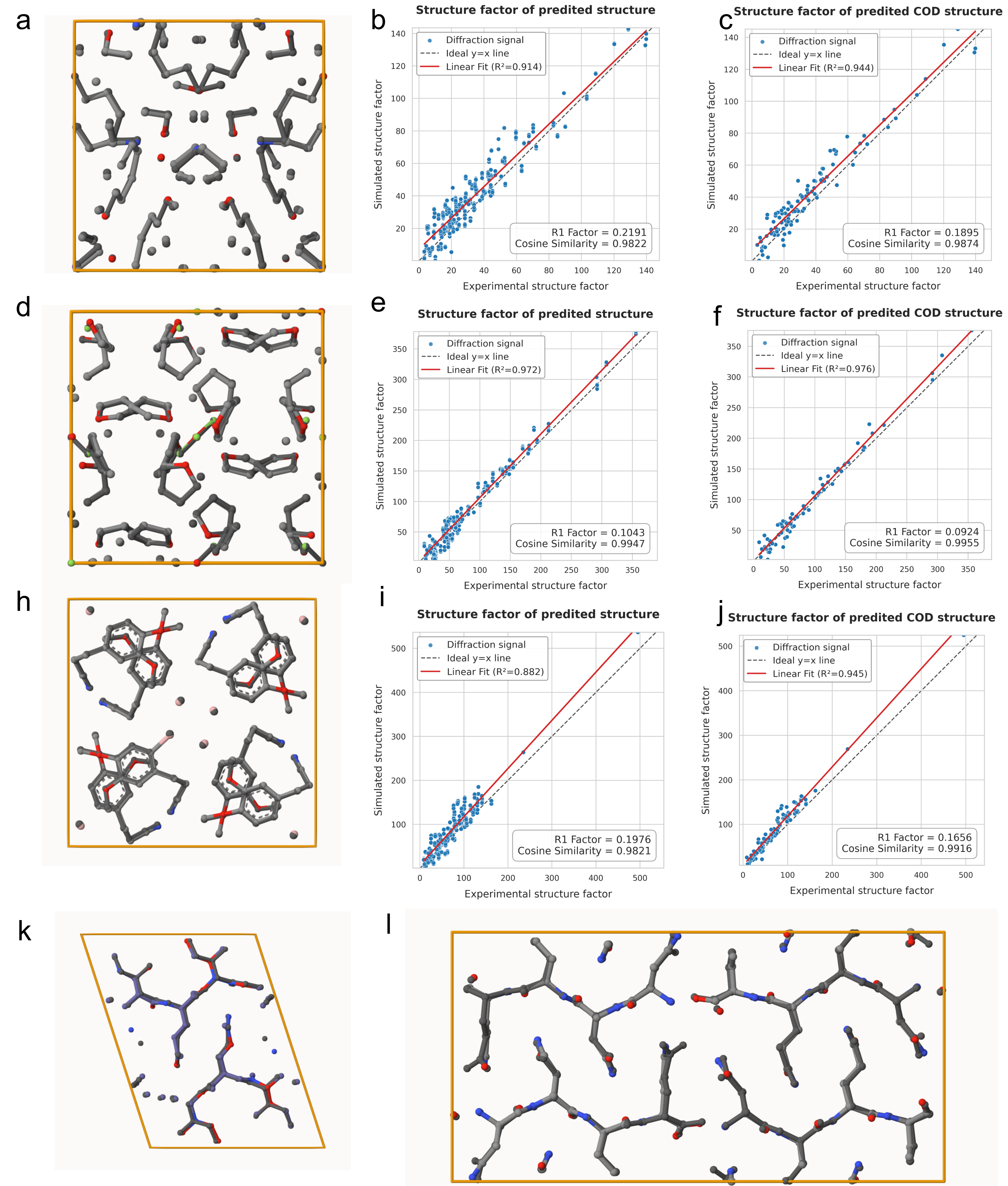}
\caption{
\textbf{XDXD predictions of small molecule and peptide structures. Predicted structures are shown in darker colors, while reference structures from the COD or RCSB are shown in lighter colors.}  
\textbf{a-c.} Small molecule structure with chemical composition C$_{12}$H$_{19}$NO$_{2}$ and space group \textit{P -4 21 c}.
\textbf{d-f.} Small molecule structure with space group \textit{P a -3} and chemical composition C$_{24}$H$_{48}$Cl$_{6}$Fe$_{3}$O$_{7}$.
\textbf{h-j.} Small molecule structure with space group \textit{P 42 b c} and chemical composition C$_{11}$H$_{12}$BrNO$_{2}$.
\textbf{b,c,e,f,i,j.} Scatter plots of observed versus simulated structure factors for the COD-retrieved structure and the predicted structure of the corresponding row.
\textbf{k.} Predicted protein structure for PDB ID 5zmz (peptide) compared with the reference structure from RCSB.
\textbf{l.} Predicted protein structure for PDB ID 2olx (peptide) compared with the reference structure from RCSB.
}\label{fig5}
\end{figure}

\textbf{Figure~\ref{fig5}a-j} presents three independent structure determination for small-molecule crystals at 2.0~\AA{} resolution. Visual comparisons between predicted structures and their ground-truth counterparts from the Crystallography Open Database (COD) are rendered in \textbf{Fig.~\ref{fig5}a, d, and h}, with reference structures shown in {dark gray}. These results demonstrate exceptional geometric congruence in atomic positioning, confirming robust structural reconstruction capability. Quantitative validation was performed through reciprocal-space analysis comparing experimental structure factors with those derived from predicted and reference structures. \textbf{Figures~\ref{fig5}b, e, and i} exhibit correlation plots for XDXD-generated structures, while parallel analyses for COD reference structures appear in \textbf{Fig.~\ref{fig5}c, f, and j}. Crucially, the R1 factors for XDXD predictions show only marginal elevation ($\Delta R_1 < 0.05$) relative to the reference structures, indicating near-equivalent fidelity to experimental diffraction data.

Notably, \textbf{Fig.~\ref{fig5}k,l} demonstrates successful oligopeptide structure determination despite complete exclusion of proteins and peptides from training data. For these biologically relevant systems, we employed extended sampling protocols (5,000 diffusion steps instead of standard 1,000) during inference on experimental peptide diffraction data. {The} oligopeptide validation dataset, including both diffraction patterns and reference atomic coordinates, were sourced from the RCSB Protein Data Bank (PDB) to ensure experimental rigor. Structural superposition analyses in \textbf{Fig.~\ref{fig5}} reveal atomic accuracy between predicted and ground-truth structures, with backbone root-mean-square deviations (RMSD) below 1.5~\AA{} across {the} validated oligopeptide. This level of precision approaches the theoretical limits of medium-resolution (2.0-3.0~\AA{}) X-ray crystallography, confirming the method's robust transferability to biologically complex systems beyond its original training domain.

\section{Discussion}

In this paper, we introduce XDXD, a pioneering end-to-end deep learning model that directly predicts crystal structures from single crystal X-ray diffraction patterns, even at the challenging resolution of 2.0 \AA{}. Trained on an extensive dataset of 395,117 simulated diffraction patterns derived from experimentally determined small molecule structures, XDXD demonstrates exceptional performance when evaluated on 24,000 real experimental structures from the Crystallography Open Database (COD). For small crystals with fewer than 52 non-hydrogen atoms, the model achieves a remarkable match rate exceeding 80\%, with a root mean square error (RMSE) lower than 0.05. {State-of-the-art crystal structure prediction methods are typically limited to systems with fewer than 52 atoms. Even for much larger systems—containing 160 to 200 non-hydrogen atoms—XDXD still achieves an impressive match rate of approximately 40\%, demonstrating its robustness and scalability.}

The success of XDXD is particularly noteworthy given the historical challenges of structure determination at low resolutions. Traditional methods, such as the Patterson method and direct methods, often struggle with low-resolution data, especially for larger systems. Molecular replacement, while useful, requires homologous models and manual refinement, which can be labor-intensive and error-prone. In contrast, XDXD offers a fully automated, data-driven approach that bypasses these limitations, providing accurate structure predictions directly from diffraction data. {Our training set comprises 80 elements—covering the vast majority of those present in the human body—while the COD test set contains 78 elements.} The model’s ability to handle structures across a wide range of space groups and elemental compositions further prove its versatility and generalizability. The evaluation on experimental data, with no overlap with the training set, confirms XDXD’s robustness and its ability to generalize to unseen structures. The high cosine similarity and low R1 values between experimental and simulated diffraction patterns validate the accuracy of the predicted structures. Furthermore, the model’s performance is consistent across various space groups and elemental compositions, as demonstrated in \textbf{Fig.~\ref{fig4}}, highlighting its broad applicability.

One of the most exciting aspects of XDXD is its potential to extend beyond small molecules to macromolecular structure determination. Although the model was trained solely on small molecule data, case studies on peptide structures (e.g., PDB IDs 5zmz and 2olx) demonstrate its capability to predict the structures of larger, biologically relevant molecules. This suggests that XDXD could serve as a foundation for future developments in macromolecular crystallography, where low-resolution data often pose significant challenges. The ability to determine macromolecular structures from limited diffraction data could accelerate research in structural biology, drug discovery, and materials science.

\backmatter
\bigskip

\bibliography{sn-bibliography}

\section{Data availability}\label{data}
All crystal structure data for training are aquired from \href{https://figshare.com/articles/dataset/Datasets\_and\_source\_data\_of\_MCRT/27844302}{MCRT}. All CIF file for real experimental evaluation are downloaded from \href{https://www.crystallography.net/cod/}{COD}.

\section{Code availability}\label{data}
The source code will be publicly accessible via GitHub upon the acceptance of our paper.

\section{Competing interests}\label{data}
The authors declare no competing interests.

\end{document}